\documentclass{llncs}

\usepackage{amsmath}
\usepackage{amssymb}
\usepackage[dvips]{graphicx}
\usepackage{color}
\usepackage{tikz}
\usetikzlibrary{matrix,shapes,arrows,positioning,chains,calc,fit}

\newtheorem{thm}{Theorem}

\begin{document}

\title{Inference Control for Privacy-Preserving Genome Matching}

\author{Florian Kerschbaum\inst{1} \and
        Martin Beck\inst{2} \and
        Dagmar Sch\"onfeld\inst{2}}
\institute{SAP Research\\ Karlsruhe, Germany \\ \email{florian.kerschbaum@sap.com} \and
           Technische Universit\"at Dresden\\ Institute of Systems Architecture\\ Dresden, Germany\\ \email{\{martin.beck1|dagmar.schoenfeld\}@tu-dresden.de}}


\maketitle

\begin{abstract}
Privacy is of the utmost importance in genomic matching.
Therefore a number of privacy-preserving protocols have been presented using secure computation.
Nevertheless, none of these protocols prevents inferences from the result.
Goodrich has shown that this resulting information is sufficient for an effective attack on genome databases.
In this paper we present an approach that can detect and mitigate such an attack on encrypted messages while still preserving the privacy of both parties.
Note that randomization, e.g.~using differential privacy, will almost certainly destroy the utility of the matching result.
We combine two known cryptographic primitives -- secure computation of the edit distance and fuzzy commitments -- in order to prevent submission of similar genome sequences.
Particularly, we contribute an efficient zero-knowledge proof that the same input has been used in both primitives.
We show that using our approach it is feasible to preserve privacy in genome matching and also detect and mitigate Goodrich's attack.
\end{abstract}





\section{Introduction}
\label{sec:intro}

Consider a situation where a client Alice wants to search her genome string in Bob's database.
Privacy is of the utmost concern in this scenario.
Neither Alice wants to reveal her genome, e.g., because it may be related to a close person, nor Bob wants to reveal his database, e.g., because of privacy legalization or intellectual property issues.
There are a number of privacy-preserving protocols based on secure computation that can reconcile this conflict, e.g.~\cite{AtaKer03,BecKer12,JhaKru08,TroKat07}.
Using secure computation (or homomorphic encryption) neither Alice nor Bob need to reveal their strings to the other party, but will learn the result of the matching.

Nevertheless, Goodrich has shown that even in the case of a privacy-preserving protocol significant information may often be leaked~\cite{Goo09}.
Using only the result of the comparison Alice can guess Bob's string by repetitive queries.
Surprisingly few queries are necessary to infer even long, real-world sized genomes.

In this paper we will design a privacy-preserving protocol that also can detect (and mitigate) whether an inference attack is taking place.
If it is not, i.e., in the case of an honest Alice, both parties' privacy will be protected and Alice will learn the result.
If an attack is taking place, Alice will not receive the answer to ``suspicious'' queries.
We use Goodrich's attack as a template, but we detect and mitigate all types of attacks that localize a specific nucleotide in the genome (character in the string).
We validate against Goodrich's attack, since it is well known and designed to be particularly efficient in this type, but our approach is more general.

Attack detection on encrypted data is a complicated, not straight-forward approach.
First, Bob may not know Alice's query string in order to preserve privacy.
It must remain encrypted.
Therefore the inference control algorithm must also work on encrypted data; to the best of our knowledge we are the first to develop and analyze such an inference control algorithm.
This could be still implemented using a generic secure computation, but, second, all inference control algorithms work on the entire history of Alice and match the query against all previous queries by Alice~\cite{Dom08}.
Therefore the secure computation must span the entire history of queries, quickly making it impractically inefficient.
Instead, in our solution the cryptographic operations are limited to a single query and inference control can be performed using simple (plaintext) equality matching of the history.
Yet, we need to trade some precision of the inference control mechanism, e.g., compared to logic-based methods, for this efficiency while still effectively mitigating the attack.

Our construction achieves this by using two different primitives and basically works as follows:
We use the secure computation genome matching protocol based on Bloom filters and homomorphic encryption of~\cite{BecKer12}.
In addition to the encrypted Bloom filter Alice submits a fuzzy (but deterministic) commitment~\cite{JueWat99} of the Bloom filter.
Bob can match those fuzzy commitments and will not answer the query if Alice has committed to a similar value before, such that ``close'' queries are prevented.
The idea is that then an uncertainty about Bob's string remains.

An obvious attack on the scheme is that Alice could use different Bloom filters in secure computation and fuzzy commitment.
So, an important contribution of the paper is an efficient zero-knowledge proof -- not resorting to generic constructions -- that Alice actually used the same Bloom filter in both.
Furthermore, Alice could compute the error-correcting code incorrectly.
For this, we also contribute a zero-knowledge proof that Alice submitted an information word.

We emphasize that our approach to mitigate Goodrich-like attacks is detective.
Our algorithm groups close patterns to match an on-going attack.
As such our inference control algorithm for mitigating attacks is probabilistic and thus may give false-positive and -negative results.
This implies that there is no provable guarantee that an attacker cannot infer the genome in the database and furthermore, some legitimate queries may be rejected.
On the contrary, our guarantee of privacy -- and soundness in the zero-knowledge proof -- is provable.

In summary, this paper contributes
\begin{itemize}

\item a privacy-preserving genome matching scheme that allows {\em detection of similar inputs}

\item an efficient {\em zero-knowledge proof} that the client has submitted its input truthfully and does not evade detection

\item an {\em analysis} of our detection scheme and how it mitigates Goodrich-like attacks on privacy of genome matching.

\end{itemize}

The remainder of the paper is structured as follows.
In Section~\ref{sec:related} we review related work on privacy-preserving genome matching.
We introduce our building blocks used in our protocol in Section~\ref{sec:blocks}.
We describe our zero-knowledge proof in Section~\ref{sec:zkp} and its security proof in Section~\ref{sec:proof}.
Section~\ref{sec:analysis} presents the analysis of our scheme under Goodrich's attack.
Section~\ref{sec:conclusion} concludes the paper.

\section{Related Work}
\label{sec:related}

Privacy-preserving matching of genomes has been introduced in~\cite{AtaKer03}.
It presents a secure computation based on homomorphic encryption, such that both the querier's genome and the database's genome are protected.
The algorithm implemented in the secure computation is edit distance.
This setup has been improved in performance using Yao's protocol~\cite{Yao86} for secure computations in~\cite{JhaKru08}.
Although further improvements to Yao's protocol yield even better performance in this computation~\cite{HuaEva11} it is still too slow for large scale deployment.

Therefore different approaches to computing the edit distance were sought.
Automata and regular expressions can emulate edit distance computations efficiently for small edit distances.
An oblivious evaluation of automata is presented in~\cite{TroKat07}, but due to the regular expressions it does not scale to real-world sized genomes.
Bloom filters can also be used to estimate the edit distance.
An evaluation of this approach using homomorphic encryption is presented in~\cite{BecKer12}.
This approach yields reasonable run-times (approx.~5 minutes) for real-world sized genomes  (approx.~16500 characters) and we therefore build upon this approach.


In order to improve performance simplified algorithms -- compared to edit distance -- are considered.
In~\cite{BalBar11} simple set intersection is used to match genomes.
Of course, the applicability to real bioinformatics is limited.
Instead, bioinformatics uses increasingly complex algorithms and some have already been made privacy-preserving.
Hidden Markov models are used to privacy-preservingly analyze gene sequences in~\cite{FraKat11}.

Although secure computation offers a formal security model -- semi-honest security~\cite{Gol04}, it does not prevent inferences from the result.
Goodrich therefore presented an attack based on the information of the edit distance alone~\cite{Goo09}.
With very few repeated queries it can infer real-world sized genomes.
The contribution of this paper is to combine the approach of~\cite{BecKer12} with mitigation of this and similar attacks.

The guaranteed randomization approach to prevent inferences about the input is differential privacy~\cite{Dwo08}.
A randomized function $K$ gives $\epsilon$-differential privacy if, for all data sets $D_1$ and $D_2$ differing on at most one element and all $S \subset Range(K)$,
$$ Pr[K(D_1) \in S] \le \exp(\epsilon) \cdot Pr[K(D_2) \in S] $$
It means, that the likelihood of any function result will only marginally change with the presence or absence of one additional element. 

Unfortunately, in our scenario the sensitivity of the function $K$ is the maximum distance between any two possible genomes, i.e., the length of the genomes themselves.
Therefore, the probability of any query result (edit distance) -- regardless of the genomes -- may change by at most a factor of $\exp(\epsilon)$.
Clearly, this completely annihilates the utility of any differentially private query result, since the result distribution quickly approaches the uniform distribution with decreasing $\epsilon$.
Furthermore, we want to protect the presence of a genome in the database against a series of edit distance queries with other genomes.
In repeated queries the security parameter $\epsilon$ adds up, since the same genome in the database may be queried~\cite{McS09}.
We therefore propose the competing approach of detecting similar queries in this paper.

Further attacks on privacy mechanisms in genomic computing have to be considered.
Bloom filter matching using the approach of~\cite{BelChe04} (keyed cryptographic hash functions instead of regular hash functions) is insecure~\cite{KuzKan11}.
A sophisticated attack can infer information from a Bloom filter with unknown hash functions using environmental information, such as the space of all genomes.
We therefore use homomorphic encryption to protect the Bloom filter and are not susceptible to this attack.
Anonymization techniques have also been found to be insecure~\cite{WanLi09}.

A related problem to privacy-preserving genome matching between two parties is outsourcing of this computation.
This has been first considered in~\cite{AtaLi04}.
A protocol for two servers executing a secure computation is presented.
The protocol of~\cite{TroKat07} has been used for outsourcing in~\cite{BlaAli10}.
The protocol of~\cite{JhaKru08} has been used in~\cite{BlaAta12}.
A clever technique of partitioning the problem into a coarse and a fine-granular part has been presented in~\cite{ChePen12}.
An approach for simple queries on an encrypted, outsourced genome database are presented in~\cite{KanJia08}.

\section{Building Blocks}
\label{sec:blocks}

\subsection{Homomorphic Encryption}
\label{sec:he}

Homomorphic encryption supports a homomorphism of (at least) one arithmetic operation on the ciphertexts to an arithmetic operation on the plaintexts.
Additively homomorphic cryptosystems support addition as the homomorphic operation on the plaintexts.
Let $E(x, r)$ denote the encryption of plaintext $x$ with randomization parameter $r$.
Then the following addition properties hold
\begin{eqnarray*}
E(x, r) E(y, s) & = & E(x + y, rs) \\
E(x, r)^y       & = & E(x y, r^y)
\end{eqnarray*}
For readability reasons we write the simplification of $E(x) = E(x,r)$, which also always uses a fresh $r$ implicitly.
Particularly we use the encryption scheme of Boneh, Goh, and Nissim~\cite{BonGoh05}.
This scheme is somewhat homomorphic, because it also supports one round of multiplication.
Let $E'(x, r)$ denote a second encryption of plaintext $x$ with randomization parameter $r$.
This scheme is constructed from a bilinear map $\hat{e}(c, d)$.
Let $\eta$ be a fixed parameter and $\theta$ a randomization parameter\footnote{If we choose the generators in the group of the result of the bilinear map cleverly, then we can compute $\eta$ and $\theta$.}, then the following multiplication property holds
\begin{displaymath}
\hat{e}(E(x, r), E(y, s)) = E'(x y, \eta + xs + yr + {\theta}rs)
\end{displaymath}
For the encryption $E'()$ the above addition properties hold again.

The Boneh-Goh-Nissim (BGN) scheme has been proven indistinguishable under chosen-plaintext attack (IND-CPA) under the subgroup decision problem in~\cite{BonGoh05}.
This indistinguishability implies that an adversary cannot distinguish a ciphertext even if it is from the same plaintext without the private key.
Somewhat homomorphic encryption is less powerful than fully homomorphic encryption which supports arbitrary operations on finite fields, but also significantly more efficient, since it does not require the error-correction operation.

The BGN scheme uses asymmetric keys and in our scenario the server has only the public key and the client also has the private key.
This means the client can perform encryption $E(x)$ and decryption $D(c)$ whereas the server can only perform encryption $E(x)$.
This implies that all ciphertexts remain indistinguishable to the server unless the plaintext is explicitly disclosed in the protocol.

\subsection{Bloom Filters}
\label{sec:bf}

A Bloom filter resembles a data structure that supports two operations: storing of arbitrary elements and checking whether a given element has already been added to the filter.
Let $b$ be a bit-array of length $l$ and $b[i]$ the $i$-th bit within this array.
Further let $h_1()$,\,\dots,$h_k()$ be $k$ hash functions with results in $[1,l]$ and a uniform output distribution.
For initialization $l$, $k$ and $h_1()$,\,\dots,$h_k()$ must be fixed.
Further set $\forall i \in [1,l]:b[i]=0$.

To insert an element $m$ into the filter, all hash functions
$h_1(m)$,\,\dots,$h_k(m)$ are evaluated and the array $b$ is set to one at the referenced positions.
Set $\forall j \in [1,k]: b[h_j(m)]=1$.

Checking the presence of an element $m'$ also includes the evaluation of all
hash functions $h_1(m')$,\,\dots,$h_k(m')$.
The element was not added to the filter in case one of the positions
$b[h_1(m')]$,\,\dots,$b[h_k(m')]$ is zero.
If all checked positions represent a one, either the element was inserted before, or one or more different elements also use these positions, giving a false-positive result.

The probability $p$ of a false-positive result for $n$ inserted elements, a
filter size of $l$ bits and $k$ hash functions is given by $$
p = \left(1-\left(1-\frac{1}{l}\right)^{kn}\right)^k
$$
where $\left(1-\frac{1}{l}\right)^{kn}$ is the probability that a single position is still zero after $n$ inserted elements.
Transposing this equation gives the calculation of the required Bloom filter length $l$ under a probability $p$ as false-positive rate and a number of elements $n$ that are to be inserted.
\begin{equation}
\label{eq:bloomLen}
 l = \frac{-1}{(1-p^{1/k})^{1/(k \cdot n)}-1} 
\end{equation}

\subsection{Genome Matching Using Bloom Filters}
\label{sec:matching}

As our main goal is searching a genome database, a primitive for comparing two genomes and finding possible matches can be used as a main building block to get to the desired functionality.
We approach the problem of comparing two genomes by first converting each into a Bloom filter.
Bloom filters, as described in section~\ref{sec:bf}, can be used to represent sets and run member tests against them.
When using Bloom filters with equal configuration, i.e. identical values for $l,k$ and functions $h_1(),\,\dots,h_k()$, set union and intersection can be performed through the use of bit-wise AND and OR operations.
Building upon this a bit-wise XOR ($\oplus$) can be used to calculate the Hamming distance as an appropriate measure to solve the similarity problem and thus closely approximate the edit distance as proposed in~\cite{BecKer12}.
For brevity, we only give an overview of the approach and refer the reader to~\cite{BecKer12} for details.
A similar analysis can be found in~\cite{DurXue12}.

The edit distance 
 consists of three basic operations applicable to single characters of a string for converting one string into the other.
These are ``replace'', ``insert'' and ``delete''.
The edit distance $ed(s_1,s_2)$ upon two strings defines the minimal number of these operations to convert string $s_1$ to $s_2$.

\subsubsection{Converting Genomes Into Bloom Filters}
\label{sec:bloomCreation}

A genome can be written as a character string by taking the nucleotide sequence derived from the single character abbreviations of all nucleotides of a single DNA strand.
Further transformations convert these representative character strings into a set of shorter grams, upon which set similarity measures allow an approximation of the original string distance measures. 

An algorithm called VGRAM~\cite{Li07} is used to convert the genome character sequence into a set of grams with variable length.
For a string $s$ a sub-string of length $q$ is called $q$-gram.
Let $s[i,j]$ be the sub-string of $s$ starting at position $i$ and ending at position $j$, thus $s[i,\left(i+q-1\right)]$ defines the $q$-gram starting at position $i$ in $s$.
As the resulting elements are independent of their original position, information about repititions within strings and permutations between sub-strings is not captured anymore by the set representation.
To accommodate this, the $q$-grams are prefixed with their original position within the genomic string.
A positional $q$-gram is then defined as $\left(i,s[i,\left(i+q-1\right)]\right)$.
Following the VGRAM algorithm, a range of possible values for $q$ is defined by $[q_{\text{min}},q_{\text{max}}]$.
Positional grams within this range are generated and checked using a predefined dictionary, constructed over a reference set of genome sequences using the ``Human Mitochondrial Genome Database''~\cite{mtDnaDb06}.
The resulting grams are chosen by the VGRAM algorithm to not be too common within the whole initial sequence.
This means larger $q$ values are used for frequent grams, whereas shorter grams are preferred if a sub-string is not well represented within the dictionary.
For the final set construction, the positional information is removed to also allow grams with insertions or deletions on previous positions and thus changed indices to be matched.
The function to run an already trained VGRAM algorithm on a string $s$ is denoted by $\text{VGRAM}(s)$.

To even out the underrepresented characters at the beginning and end of the sequence, an extension of $q_{\text{max}}-1$ identical characters, which are not part of the original alphabet, is appended on both sides. Positional $q$-grams on extended strings were introduced by Gravano et al.~\cite{Gra01}.

The generated grams are treated as elements to be added to a Bloom filter.
For this, all $k$ hashes are calculated upon the grams.
As Bloom filters can be used for membership tests and estimations of set
cardinality, \cite{PapSib10} analyzed how the number of hash functions
influences the estimated results depending on the Bloom filter usage.
It was concluded, that the optimal number of hash functions to do cardinality
estimation is one.
Based on this -- we are going to perform set cardinality estimation upon the Bloom filters -- we fix $k=1$ and only use a single hash function to build and query Bloom filters throughout the rest of the paper.

\subsubsection{Encrypted Bloom Filter Similarity}
\label{sec:encBloom}

\begin{figure}
\centering
\begin{tikzpicture}
\matrix (m)[matrix of nodes, column sep=0.5cm,row  sep=8mm, nodes={draw=none, anchor=center,text depth=0pt} ]{
Alice on input $s_A$ & & Bob on input $s_B$\\[-4mm]
Generate grams & & Generate grams \\[-7mm]
$g_A = \text{VGRAM}(s_A)$ & & $g_B = \text{VGRAM}(s_B)$ \\[-7mm]
Build Bloom filter & & Build Bloom filter \\[-7mm]
$b_A[h(g_A[i])] = 1$ & & $b_B[h(g_B[i])] = 1$ \\[-7mm]
Encrypt Bloom filter & & \\[-7mm]
$a[i] = E(b_A[i], r)$ & Send $a$ & Encrypted XOR\\[-7mm]
 & & $E(b_A[i] \oplus b_B[i]) =$\\[-7mm]
 & & $E(b_B[i]) \cdot a[i]^{2b_B[i]-1}$ \\[-7mm]
 & & Encrypted distance\\[-7mm]
 & & $E(d_h(b_A,b_B)) =$ \\[-7mm]
Decrypt and output & Send $E(d_h(b_A,b_B))$ & $\prod_{i=1}^{l}E(b_A[i] \oplus b_B[i])$ \\[-7mm]
$d_h(b_A,b_B)$ & & \\
};

\draw[shorten <=-0.5cm,shorten >=-0.5cm] (m-1-1.south east)--(m-1-1.south west);
\draw[shorten <=-0.5cm,shorten >=-0.5cm] (m-1-3.south east)--(m-1-3.south west);
\draw[shorten <=-0.5cm,shorten >=-0.5cm,-latex] (m-7-2.south west)--(m-7-2.south east);
\draw[shorten <=-0.2cm,shorten >=-0.2cm,-latex] (m-12-2.south east)--(m-12-2.south west);
\end{tikzpicture}
\caption{Protocol for privacy-preserving distance measure using Bloom filters and homomorphic encryption. Alice and Bob input their genomic strings $s_A$ and $s_B$. Alice outputs a privacy-preservingly calculated set cardinality estimation as distance measure.}
\label{fig:bloomEncProt}
\end{figure}
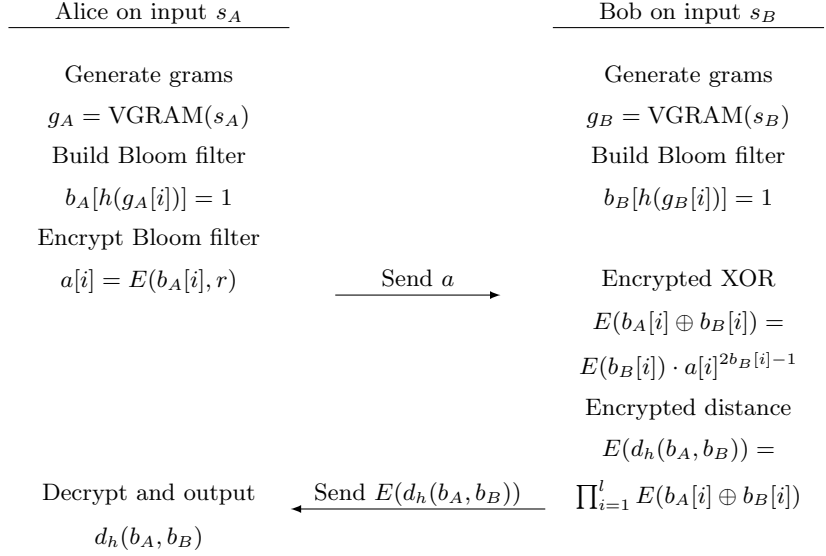

Alice wants to query Bob's database using her constructed Bloom filter $b_A$, having length $l$, as described in section \ref{sec:bf}. The protocol is shown in figure \ref{fig:bloomEncProt}.
In order to preserve privacy we use the homomorphism property of the Boneh-Goh-Nissim encryption scheme to hide the filter content.
$E(x, r)$ denotes the encryption as specified in section \ref{sec:he}.

Let $a$ be an array of size $l$ and $a[i]$ the value of the element at position $i \in [1,l]$.
Set $a[i] = E(b_A[i], r)$ for each $i \in [1,l]$.
We call $a$ the encrypted Bloom filter. 

Alice sends $a$ to Bob, who also created a Bloom filter $b_B$ for his genomic string in advance using the same number of hash functions $k = 1$, hash function $h()$ and Bloom filter length $l$. 

In~\cite{Li07} the relation between the Hamming distance of bit strings built from $q$-grams and the edit distance on character strings is formalized.
\cite{BecKer12} evaluated and experimentally showed the accuracy of approximating the edit distance using the Hamming distance between Bloom filters.
As a direct set operations are not possible anymore due to the encrypted Bloom filter $a$, we perform the Hamming distance calculation through homomorphic operations.
Keep in mind that counting all elements, which are set to one in the Bloom filter estimates the set cardinality with a low error rate, proportional to the false-positive probability of the underlying Bloom filter.
The encryption scheme allows homomorphic additions and the underlying plaintext values are all either zero or one.
This gives us the encrypted cardinality $E(|b_A|)$ of $a$ as
$$
E(|b_A|) = \prod_{i=1}^{l}a[i]
$$

However, to arrive at the Hamming distance $d_h(b_A, b_B)$ we need to calculate the XOR cardinality $|b_A \oplus b_B|$.
The encrypted XOR result between an encrypted value $E(b_A[i])$ and a plaintext value $b_B[i]$ can be defined as:
$$
E(b_A[i] \oplus b_B[i])=\begin{cases}
E(b_A[i])& \text{if $b_B[i] = 0$},\\
E(1) \cdot E(b_A[i])^{-1}& \text{if $b_B[i] = 1$}.
\end{cases}
$$


Bob will not learn the values $E(b_B[i])$ and $E(b_A[i] \oplus b_B[i])$, as they are encrypted.
Based on these Bob can calculate the encrypted Hamming distance $E(d_h(b_A,b_B))$ as follows:
$$
E(d_h(b_A,b_B)) = \prod_{i=1}^{l}E(b_A[i] \oplus b_B[i])
$$
Once the encrypted Hamming distance is sent back to Alice, it can be decrypted using her private key.

This protocol follows the approach from~\cite{BecKer12} with the modification of using a false-positive probability of $p=0.5$ for the Bloom filter.
As a result the generated Bloom filters $b_i$ with length $l_i$ have a Hamming weight of rougly $w_h(b_i) = l_i/2$ and thus $w_h(b_i)/l_i \approx 0.5$.
Particularly the calculated values of $w_h(b_i)/l_i$ for $10000$ samples have an average of $0.484$ with a standard deviation of $0.0059$.
This is important for the employed error-correcting code, as it needs an rougly equal amount of zeroes and ones for proper correction towards the codewords.
Another positive effect is the shortened Bloom filter length $l$ compared to the original work in~\cite{BecKer12}.
For full mitochondrial DNA sequences with approximately $16500$ nucleotides, a Bloom filter length of $23905$\,bit is used instead of the originally proposed $157261$\,bit, as can be derived from equation \eqref{eq:bloomLen}.
In return the number of homomorphic operations on the client and server side is greatly reduced.
The mean time for comparing a full sized mitochondrial genome is therefore lowered by $77\%$ compared to~\cite{BecKer12} for an otherwise identical configuration.
The accuracy of approximating the edit distance if not affected, as the pearson correlation between the edit distance of the original strings and our distance measure is still at $0.997$.

\subsection{Error-Correcting Codes}
\label{sec:ecc}

Error-correcting codes are commonly used to correct errors during transmission.
Let $x$ be a bit-string of length $l'$ bits.
We call $x$ the information word. 
An error-correcting code will then transform the information word into a codeword of length $l$ bits with ($l - l'$) check bits.
The resulting codeword $y$ is then actually transmitted by Alice.
The transmission might be faulty and Bob receives $y'$.
On the receiving end Bob can recover $x$ from $y'$ as long as the distance between $y$ and $y'$ is below a certain threshold.
The performance of a code and, therefore, the threshold $t$ is described by the minimum Hamming distance $d_{\text{min}}$, defined as the minimum distance among all possible distinct pairs of $2^{l'}$ codewords in the code alphabet.
The maximum number of correctable errors can be calculated with $t =\lfloor(d_{\text{min}} - 1) / 2\rfloor$.
Of course, a high performance $t$ requires a high amount of check bits ($l - l'$).

In this paper the investigations are based on $(l,l',d_{\text{min}})=(2^m,\sum\limits_{i=0}^1{m\choose i},2^{m-1})$ first-order Reed-Muller codes~\cite{Mul54,Ree54}.
Furthermore, we use a first-order code, such that we only have first-order decoding equations, i.e., each decoding equation sums two and only two received bits.
Most importantly, we use {\em modified} Reed-Muller codes.
We modify the code, such that it is shortened and has an odd number of decoding equations for the information word bit as far as this is possible.
This has the purpose that each word $y'$ has exactly one associated information word $x$.
In particular, we avoid by this construction decoding failures where the decoding equations result in an equal number of $0$s and $1$s.
In this paper the parameters are $m = 15$ and $l < 2^m$ dependent on the length of the Bloom filters.

\subsection{Fuzzy Commitment}
\label{sec:fc}

A commitment scheme binds the committer to a certain value (binding property) without revealing its value (security property).
A fuzzy commitment scheme binds the committer to a value within a distance of the committed value.
The distance metric can again be the Hamming distance.
The fuzzy commitment scheme of~\cite{JueWat99} commits to a randomly selected codeword $y$ and secret key $\kappa$ with $MAC(y, \kappa)$ and adds a distance sequence $\tau$, such that the committed value $v = y \oplus \tau$.
When testing whether a value $z$ is covered by the commitment, one first computes $z' = z \oplus \tau$ and then performs error-correction to $z''$ of $z'$.
If $MAC(z'', \kappa) = MAC(y, \kappa)$, then $z$ is covered by the fuzzy commitment of $v$.

This type of fuzzy commitment is perfectly suitable for privacy-preserving storage of biometrics, but in our application we require the element in the commitment to be checkable (and not random).
More similar to~\cite{DavFra98}, we treat the value to be committed as the $l$-bit-string $y'$.
First, $y'$ is decoded to the information word $x$ of length $l'$ and $x$ is committed using a MAC function ($MAC(x, \kappa)$).
To check whether a $l$-bit-string $y''$ is covered by the commitment to $x$, one performs decoding to the information word $x''$ and then checks whether $MAC(x, \kappa) = MAC(x'', \kappa)$.


In our construction Alice will commit to her genome sequence using this commitment scheme with a pre-defined MAC function (see Section~\ref{sec:mac}), i.e., she sends $MAC(x, \kappa)$ along with the encrypted Bloom filter.
Now, if Alice submits two Bloom filters for genome strings within a certain distance, she will likely commit to the same value $x$ using the fuzzy commitment scheme.
Bob will be able to detect this and can withhold his answer.

Note that in our construction of the fuzzy commitment using modified, first-order Reed-Muller codes it is possible that two committed values $y'$ and $y''$ are close, but are decoded into different information words and consequently fuzzy commitments.
This is unavoidable except in a perfect code which only exists for $t = 1$ or $t = 3$, but results in a false-negative of the fuzzy commitment.
We empirically estimate this error and in our experiments for codes with length $l = 23905$ ($m = 15$) using equation \eqref{eq:bloomLen}.
The probability for two $l$-bit-strings $y'$ and $y''$ within Hamming distance $\delta = d_h(y', y'') \leq 20$ to decode into two different information words is less than $10\%$.
Furthermore, this probability is decreasing with increasing length $l$.


\section{Zero-Knowledge Proof}
\label{sec:zkp}

In this section we describe the zero-knowledge proof (ZKP) that the client Alice submitted the same Bloom filter in the homomorphic encryption and (corresponding) information word in the fuzzy commitment scheme.
Obviously, the generic technique for ZKPs based on Karp reductions to languages in NP is incredibly inefficient.
We therefore provide a specialized proof for our problem based mostly on somewhat homomorphic encryption.
Our choices of error-correcting code (Section~\ref{sec:ecc}) and MAC function (Section~\ref{sec:mac}) have been specifically tailored to our choice of homomorphic encryption scheme, i.e., they are computable in the efficient scheme of Boneh-Goh-Nissim (Section~\ref{sec:he}).

Still, it is quite difficult to prove that an information word has been correctly decoded from a Bloom filter using homomorphic encryption.
In Reed-Muller codes this requires $l'$ majority (range) proofs.
Instead, we first compute the codeword $d$ for the Bloom filter, prove that it has the right distance from the Bloom filter (Section~\ref{sec:zkphamming}) and decodes to the information word $x$ (Section~\ref{sec:zkpcode}).
This requires only one range proof and several proofs of plaintexts zero and is therefore significantly more efficient.
Finally, we prove that the information word has been used in the fuzzy commitment (Section~\ref{sec:zkpmac}).
Note that neither the codeword nor the information word is seen in the clear by the database, but its computation can be performed in the clear by the client without homomorphic encryption.

In our construction we will use ZKPs that a ciphertext is an encryption of either of two plaintexts from~\cite{DamJur01} and that a vector of ciphertext is a shuffle of another vector ciphertexts~\cite{Gro10}.
Furthermore, we use a simple technique to prove that a plaintext $x$ is encrypted by ciphertext $E(x, r)$ as described in section~\ref{sec:he}:
The prover reveals the random parameter $r$.

First in the next section, we will fix the MAC function we use in the fuzzy commitment scheme.
Then in the subsequent sections, we will show how to prove input equality between encrypted Bloom filter and MAC function.

\subsection{MAC Function}
\label{sec:mac}

We replace the use of a standard MAC function in fuzzy commitments by a one-way function, as the strong security properties of forgery resistance or pseudo-randomness are not required. The one-wayness property is sufficient, since we sample from a sufficiently large, but restricted domain of $l'$-bit information words\footnote{Alice only needs to decrypt values smaller than the logarithm of the size of the domain.}.
Usually symmetric cryptography is used for speed, but the time-critical operation in our case is the ZKP.
Therefore we use a one-way function whose properties can be easily proven.

We use the homomorphic encryption function with a fixed random parameter $\kappa$.
Using a domain with more than $l'$ bits in the homomorphic encryption function we are certain to avoid collisions.
If the information word has more bits than the domain, we can use multiple encryption functions.
Then our one-way (``MAC'') function is
$$
OWF(x, \kappa) = E(x, \kappa)
$$

The key $\kappa$ is at Alice's discretion, but she has to commit to it by sending a random value $s$ and $\hat{s} = E(s, \kappa)$ to Bob during setup.
Of course, Alice has to also prove proper construction of the homomorphic encryption scheme.
Since the key is a composite of two primes, she can use another zero-knowledge proof, e.g.~\cite{CamMic99}, or for efficiency reveal the primes to an authorized third party.
Alice, also needs to prove that the generators of the encryption scheme where chosen randomly.
She can do this in a white-box way by revealing the randomness used to compute them.
Note that all of these steps only need to be completed once during setup and not for each genome Alice is querying.


\subsection{Proof of Hamming Distance}
\label{sec:zkphamming}

Alice intends to prove that the same Bloom filter was used in the homomorphic encryption and in the fuzzy commitment scheme, but the information word in the fuzzy commitment scheme is based on an error-corrected codeword.
Therefore there may be a difference between the Bloom filter and the codeword, but the difference must be small.
Alice therefore proves that the Hamming distance between the two strings is at most our chosen parameter $\delta$.

Let $b_A[i]$ ($0 \leq i < l$) be the $i$-th bit of the Bloom filter $b_A$ in the homomorphic encryption.
Let $r[i]$ be fresh random values and $a[i] = E(b_A[i], r[i])$.
Alice sends all $a[i]$ to Bob.

Alice first creates $l$ bits $d[i]$, such that $d$ is the error-corrected codeword of $b_A$.
Then $d_h(b_A, d) \leq \delta$.
Let $r'[i]$ be fresh random values and $c[i] = E(d[i], r'[i])$.
Alice sends all $c[i]$ to Bob.

Alice proves in zero-knowledge that all $a[i]$ and $c[i]$ are indeed encrypted bits, i.e., their plaintext is either $0$ or $1$.
She uses the ZKP from~\cite{DamJur01}.

Alice and Bob now compute
$$
E'(d_h(b_A, d), r) = \prod_{i = 0}^{l} \hat{e}(a[i] c[i]^{-1}, a[i] c[i]^{-1})
$$
They compute for each $0 \leq j \leq \delta$:
$$
f[j] = E'(g[j], r') = E'(d_h(b_A, d), r) E'(-j, 0))
$$
Note that for $j = d_h(b_A, d)$ it holds that $g[j] = 0$ and $g[j] \neq 0$ otherwise.

Alice performs a random shuffle of all $f[j]$.
Let $f[{\pi(j)}]$ denote the elements in this shuffle.
Alice sends the shuffled $f[{\pi(j)}]$ and proves in zero-knowledge that it indeed is a shuffle.
Alice then reveals $f[{\pi(d_h(b_A, d))}] = E'(0, r'')$ and $r''$ to Bob.
Bob verifies that $f[{\pi(d_h(b_A, d))}]$ is in the shuffle and an encryption of zero (with the revealed random parameter $r''$).

Bob now knows that $d$ is a bit-string within Hamming distance $\delta$ of Bloom filter $b_A$.

\subsection{Proof of Code and Information Word}
\label{sec:zkpcode}

In order to prove that $d$ is a codeword with information word $x$ Alice needs to prove that all decoding equations have the same result for each bit.
This follows from the construction of Reed-Muller codes~\cite{Mul54,Ree54}.
The generator matrix and therefore the equations are known to both -- Alice and Bob.
We operate $\left(\textrm{mod} \enspace 2 \right)$, i.e., all operations in the equations are XOR.

Alice and Bob build all equations for each information word bit $x[i]$.
Let $x'[j]$ ($0 < j < 2^{m}/2$) be the result of the $j$-th equation for $x[i]$.
Alice and Bob can compute the result $E(x'[j])$ from $E(d)$ using the homomorphic operation.
Recall that we use modified Reed-Muller codes and further utilize only first-order equations, i.e. two decoding bits.
Let $d[j']$ and $d[j'']$ be the two bits decoding to $x'[j] = d[j'] \oplus d[j'']$.
Then
$$
\setlength{\arraycolsep}{0pt}
\begin{array}{rl}
E'(x'[j], r'[j]) = \hat{e}( & E(d[j'], r') E(d[j''], r''), E(1, 0)) \\
                   \hat{e}( & E(d[j'], r'), E(d[j''], r''))^{-1}
\end{array}
\setlength{\arraycolsep}{3pt}
$$

Alice sends also the bit-wise encrypted information word $E(x[i], r[i])$ ($0 \leq i < l'$) to Bob.
Alice and Bob compute the second ciphertext for each bit $E'(x[i], r'''[i]) = \hat{e}(E(x[i], r[i]), E(1, 0))$.
Then, Alice and Bob compute the differences between the bit information word and each decoding result (in one variable)
\begin{eqnarray*}
E'(x''[i], r''[i]) & = & \prod_{j = 1}^{2^{m}/2 - 1} (E'(x[i], r'''[i])^{-1} E'(x'[j], r'[j]))^{2^j} \\
                   & = & E'(\sum_{j = 1}^{2^{m}/2 - 1} 2^j (x'[j] - x[i]), r''[i])
\end{eqnarray*}
Now, this difference must always be zero and Alice proves that all $x''[i] = 0$ by revealing all $r''[i]$.
Bob now knows that $x$ is the information word for codeword $d$.

\subsection{Proof of OWF Computation}
\label{sec:zkpmac}

Alice sends $OWF(x, \kappa) = E(x, \kappa)$ to Bob, which is referred to as $E(x', \kappa)$.
Alice has already proven that $x$ is the information word for a codeword $d$ within Hamming distance $\delta$ of $b_A$.
She now needs to prove that the OWF was computed on $x$.
Particularly, note that Alice could cheat on using the correct key $\kappa$.

Alice and Bob compute
$$
E(x, r) = \prod_{i = 0}^{l'-1} E(x[i], r[i])^{2^{i}}
$$
and
$$
E(h, r') = E(x, r) E(x', \kappa)^{-1}
$$
Alice proves that $h = 0$ by sending $r'$.
Bob verifies that indeed $E(h, r') = E(0, r')$.

Recall that $\hat{s} = E(s, \kappa)$ is Alice's commitment to her OWF key $\kappa$.
Alice needs to prove that $E(x, \kappa)$ was also computed with the random parameter $\kappa$.
Alice and Bob compute
$$
E(u, v) = OWF(x, \kappa) E(s, \kappa)^{-1}
$$

If the OWF was computed properly, then $v = 0$ (but $u$ is randomly distributed).
Alice now needs to prove $v = 0$, but without revealing the difference of $d$ and $s$.
She therefore cannot reveal $u$.

Instead, she chooses a random value $r''$ and computes $w = E(r'', 0)$.
She sends $w$ to Bob.
Bob flips a coin $\eta \in \{ 0, 1 \}$ and either challenges Alice to reveal $r''$ (if $\eta = 0$) or $r'' + u$ (if $\eta = 1$).
Bob verifies that $w = E(r'', 0)$ (in case of $\eta = 0$) or $w E(u, v) = E(r'' + u, 0)$.
This is repeated $\lambda$ times.
Alice's chance of successfully cheating is $2^{-\lambda}$.

\section{Security Proof}
\label{sec:proof}

\begin{thm}
Our ZKP is complete, sound and honest-verifier zero-knowledge.
\end{thm}

Our ZKP is a composition of several smaller ZKPs.
We therefore either prove their completeness, soundness and zero-knowledge properties first or reference the relevant literature.

We begin in reverse order with the ZKP whether $v = 0$ in $c = E(u, v)$.
It is {\em complete}, because if $v = 0$, then in both options -- $E(r'', 0)$ and $E(r'' + u, 0)$ -- the randomization parameter is $0$.
It is {\em sound}, because if $v \neq 0$, then either in $E(r'', v')$ or $E(r'' + u, v'')$ the randomization parameter -- $v'$ or $v''$ is not $0$.
The prover will be caught with probability $\frac{1}{2}$.
It is {\em honest-verifier zero-knowledge}, because the following simulator does not require the secret input, i.e., the value $u$ or the secret key.
First, the simulator outputs an uniformly chosen $\rho$ as the last message.
If the verifier selects $\eta = 0$, it sends $E(\rho, 0)$ as the first message.
If the verifier selects $\eta = 1$, it sends $w = E(\rho, 0) E(u, v)^{-1}$ as the first message.

Next, we consider the ZKP whether $x = 0$ in $c = E(x, v)$.
The completeness, soundness and zero-knowledge property of this ZKP were given in~\cite{DamJur01}.
Based on this, the proofs for one of two plaintexts were also given in~\cite{DamJur01}.
Furthermore, the completeness, soundness and zero-knowledge property of the shuffle ZKP were given in~\cite{Gro10}.

We now need to prove the completeness, soundness and zero-knowledge property of the composite ZKP.
For brevity we do not prove each step individually, but only present the summary of our construction.
We follow the composition theorem for semi-honest secure computations from Goldreich~\cite{Gol04} and replace the sub-ZKP by oracle functionality.
The composite ZKP is {\em complete}, because if the fuzzy commitment $OWF(x, \kappa)$ and the encrypted Bloom filter $a[j]$ ($0 \leq j < l$) relate to the same information word $x$, all oracle ZKPs will succeed.
The composite ZKP is {\em sound}, because if the fuzzy commitment $OWF(x, \kappa)$ and the encrypted Bloom filter $a[j]$ use non-related information words, one oracle ZKPs will not succeed.
This is an abbreviation, since we would need to show how the prover could deviate and then be caught.
Instead -- for brevity -- we leave this as a proposition.

In order to show honest zero-knowledge we give the following simulator.
First, Alice sends the encrypted Bloom filter $a[j]$ and $c[j]$ ($0 \leq j < l$).
These can be simulated using random plaintexts and randomization parameters, since the encryption is IND-CPA secure~\cite{BonGoh05}.
Now, we invoke the simulator for ZKP that the plaintext is one of either two from~\cite{DamJur01}.
Note that this simulator will be correct, since it is independent of the secret input.
The subsequent computations are performed by both -- Alice and Bob.
We then choose $\delta + 1$ random ciphertexts including $E'(0, r'')$.
We invoke the simulator for the shuffle from~\cite{Gro10}.
Again, recall that the simulator is independent of the secret input.
Furthermore, due to the IND-CPA security our simulated shuffle is indistinguishable from a real shuffle.
The simulator reveals the random parameter $r''$.


The simulator computes the information word $x$ for the (random) codeword $d$ ($c[j] = E(d[j], r'[j])$ for $0 \leq j < l$).
It chooses random $r[i]$ ($0 \leq i < l'$) and outputs the ciphertexts $E(x[i], r[i])$ for the information word.
Again, these are IND-CPA indistinguishable.
Furthermore, it performs the computations as in Section~\ref{sec:zkpcode} and outputs the corresponding $r''[i]$.

The simulator chooses random $r'$ and sends from the previous simulated messages $E(x, \kappa) = \prod_{i = 0}^{l'-1} E(x[i], r'[i])^{2^{i}} E(0, r')^{-1}$ as the fuzzy commitment.
It is indistinguishable again due to the IND-CPA security of the encryption scheme.
The subsequent computations are again performed by both -- Alice and Bob.
The simulator reveals $r$.
The simulator invokes the simulator for the ZKP whether $v = 0$ in $c = E(u, v)$ from above.
Again, this simulator succeeds, since it is independent of the secret input.

We have given a complete simulator -- independent of the secret input by Alice -- for the message exchange between prover and verifier.
It uses the simulators of the oracle ZKPs as subcomponents.

\section{Analysis of Goodrich's Attack}
\label{sec:analysis}

Goodrich~\cite{Goo09} describes two attacks to discover private strings using similarity scores obtained in a privacy-preserving way.
One of the two attacks Goodrich describes relates to the black-peg score of the Mastermind game and can easily be adopted to be used against edit distance scores.
Again, for brevity we only give an overview of the attack and refer the reader to~\cite{Goo09} for details.

\subsection{Description of the Original Attack}
\label{sec:goodDescr}

The generic algorithm starts with a reference string $r_0$
and an alphabet $A$, containing $j = |A|$ characters to
generate $j-1$ strings $\{r_1,\,\ldots,r_{j-1}\}$. These strings are generated such that the set of characters over all strings $r_i$ at a fixed position equals the alphabet $A$. The additional strings $r_i$ $(1 \leq i < j )$ are deviations from $r_0$, which defines the base
character for a specific position. For each string $r_i$ $(1 \leq i < j)$, the character for a position is calculated by taking the base character in
$r_0$ at that position and go $i$ characters further in alphabet $A$.

For all strings $r_i$ $(0 \leq i < j)$ the $j$ similarity scores are obtained and
the algorithm goes on recursively by dividing the reference string into two equal parts, while
again generating $j-1$ strings for the left part and query those $j-1$ scores.
The scores for the right part can be calculated using the left-part scores and
the knowledge from the previous queries.

If at some point the similarity score equals zero for one query, then the used
deviation from the original character set can be ignored for deeper recursive
steps and the number of strings to be generated and queried reduces by one. At
some point the number of queried strings drops below two and the recursion
stops.


\begin{figure*}
        \centering
        \begin{minipage}[b]{0.45\textwidth}
                \centering
                \includegraphics[width=\textwidth]{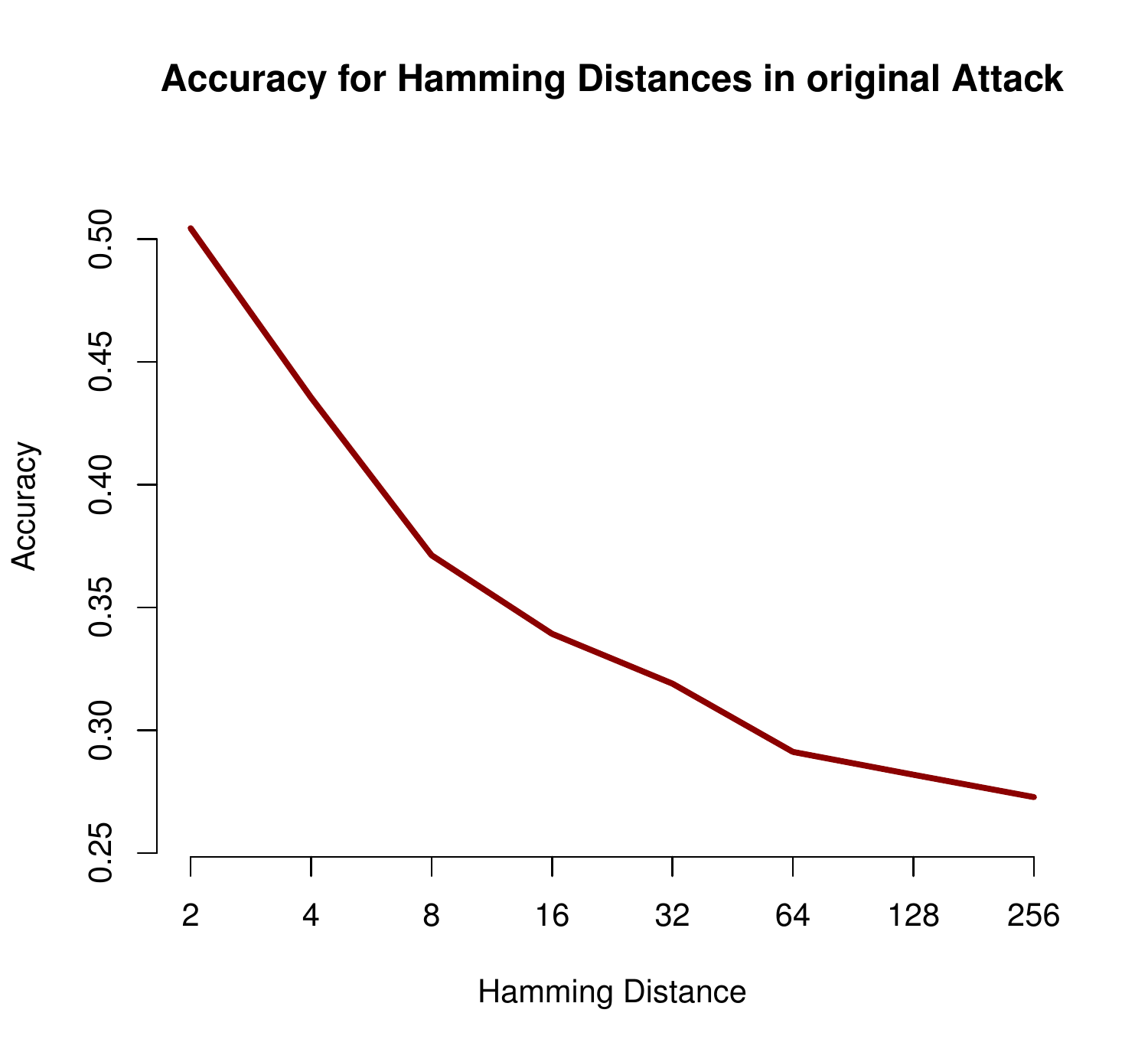}
                \caption{Accuracy of original Goodrich attack}
                \label{fig:origAttack}
        \end{minipage}%
        \hspace{12pt} 
        \begin{minipage}[b]{0.45\textwidth}
                \centering
                \includegraphics[width=\textwidth]{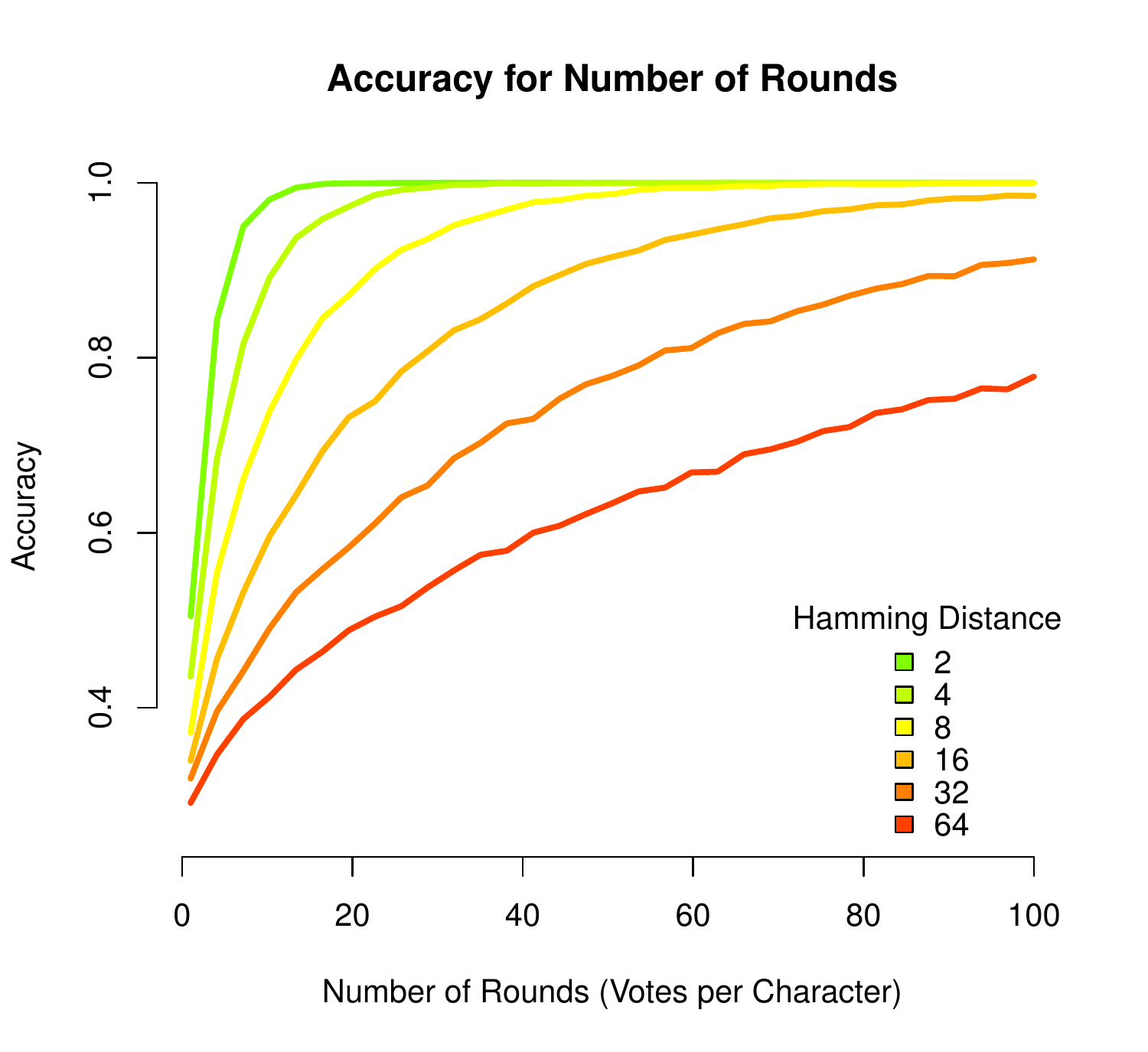}
                \caption{Accuracy after randomization rounds}
                \label{fig:randAttack}
        \end{minipage}
\end{figure*}

\subsection{Goodrich Randomization}
\label{sec:goodRand}

\begin{figure*}
        \centering
        \begin{minipage}[b]{0.45\textwidth}
                \centering
                \includegraphics[width=\textwidth]{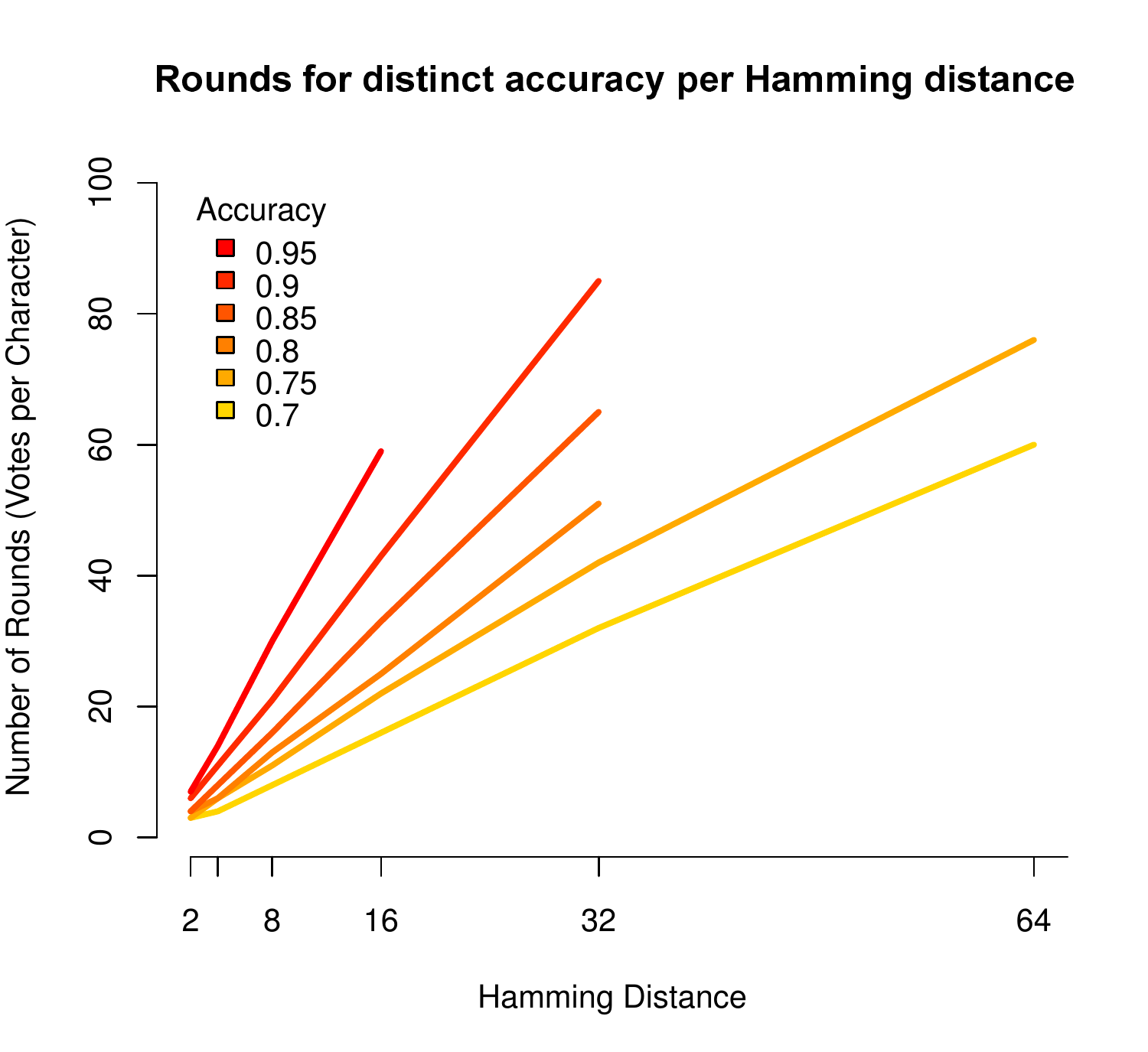}
                \caption{Rounds necessary to reach distinct accuracy}
                \label{fig:accuracyPlateus}
        \end{minipage}%
        \hspace{12pt} 
        \begin{minipage}[b]{0.45\textwidth}
                \centering
                \includegraphics[width=\textwidth]{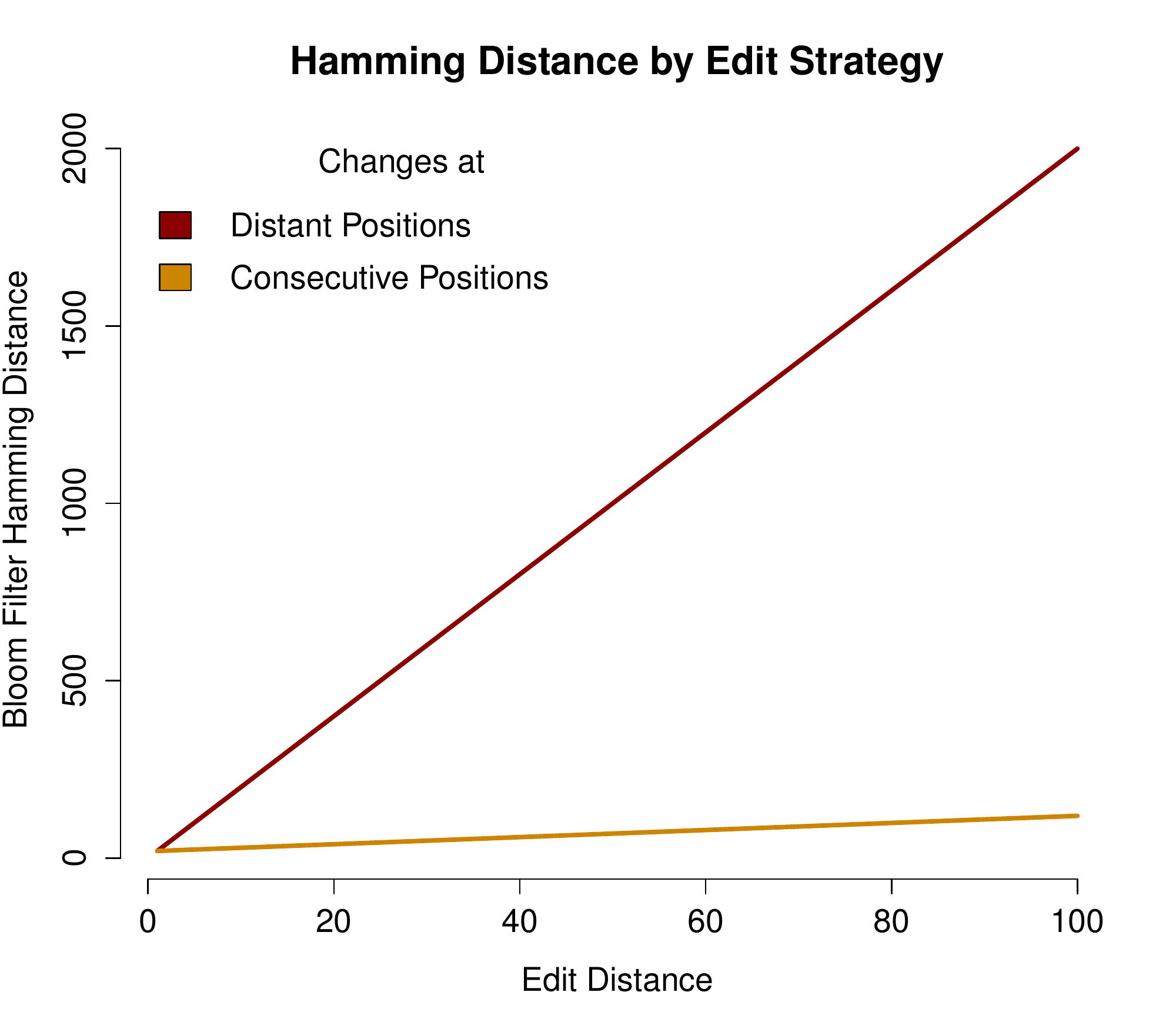}
                \caption{Bloom filter changes by distance of edit positions}
                \label{fig:bloomDistance}
        \end{minipage}
\end{figure*}

We presented in~\ref{sec:fc} the use of an error-correcting code, specifically
a Reed-Muller code as given in section~\ref{sec:ecc}, to generate a fuzzy commitment, which can be
checked to detect queries which are covered by the same commitment, which in
turn we call similar.

Depending on the configuration of the used error-correcting code, two submitted Bloom filters within a certain
Hamming distance $\delta$ can be detected as similar and no distance score will be calculated.
As Bloom filters are set representations of the original strings, let $\delta'$ be the equivalent
approximated Hamming distance between compared sequences. The Hamming distance on strings of
equal length equals the number of positions on which the strings differ.

The Hamming distance in the following figures describes different configurations of the necessary Hamming distance $\delta'$ between two submitted queries.
If a query was generated for a sequence with less than the necessary distance, it is detected as being too close to a previous submission and the answer will be withhold.
Figure~\ref{fig:origAttack} shows how the accuracy of the original attack goes down as the nesessary distance $\delta'$ increases.
The accuracy was calculated by comparing the proposed string by the Mastermind algorithm against the correct one.
The ratio of the number of correctly extracted nucleotides over the total length of the sequence (approx. 16500 characters) is given as the accuracy value.

Of course, an attacker is not bound to the original Goodrich attack.
We present here an adapted version trying to circumvent our detection mechanism.
Instead of running the Goodrich attack only once, an attack can start with
random reference sequences and interleave the results from multiple attack runs.
The overall attack is stopped after a fixed number of rounds. After each round
all votes for the characters are remembered and after the last round a sequence
consisting of the characters with the most votes is constructed. To calculate
the accuracy score the number of similar characters between the generated
sequence and the correct sequence is divided by the length of the sequence.
Figure~\ref{fig:randAttack} shows curves for different necessary Hamming
distances $\delta'$. 
Figure~\ref{fig:accuracyPlateus} gives the linear relation between the necessary Hamming
distance $\delta'$. 

As a countermeasure to this attack we propose to also limit the number of total queries to a level dependent on the target accuracy.
For example, Goodrich reports that for $90\%$ of the genomes less than $875$ queries are needed.
Given a target accuracy level of $75\%$ with a minimum edit distance of $32$ we can allow $40$ rounds of attack, roughly $35000$ queries.
This is a significant step forward compared to the maximum number under Goodrich's attack.

\subsection{Distinguishing attacks from valid requests}
\label{sec:attackSeparation}

We should ensure that our detection algorithm does not reject legitimate queries, because their are close.
When comparing genomic sequences against a reference sequence, it yields a low
number of substitutions and even less insertions and deletions on average.
\cite{Goo09} gives an average number of $28$ substitutions for a sample set of
$1000$ mitochondrial genomic sequences. 


The changes in the genomic sequences --~if compared to the reference sequence~-- are distributed over specific regions, where mutations are statistically possible.
The artificially generated strings from the attack however show certain patterns.
This difference in how the changes are distributed is also reflected in the Bloom filter.

Recall the generation of positional grams, as it was described in section~\ref{sec:bloomCreation}.
The VGRAM algorithm~\cite{Li07} also defines the generation of NAG-vectors, which describe the ``number of affected grams'' by changing a certain amount of characters.
Based on this~\cite{BecKer12} specifies the maximum approximated Hamming distance over Bloom filters based on the NAG-vectors.
The number is corrected using the false-positive probability of the Bloom filter configuration, which in our case results in an average Hamming distance of $20$\,bits.

If a single character at a random position in the sequence is changed, the Bloom filter will have on average approximately $20$\,bit distance to the original string.
This is true as long as the edit positions are at least $q_{\text{max}}$ characters away from each other.
After $u$ changes in the sequence, we see about $20 \cdot u$ changes in the Bloom filter. This is true for nearly all genomic sequences.

However, if the $u$ changes appear consecutively, as it is with the Mastermind attack, $q + (u-1)$ changes can be seen in the Bloom filter.
This is due to the affected grams for consecutive edit operations. 
The curves showing this relation on exemplary values are given in Figure~\ref{fig:bloomDistance}.
For this example we used the proposed values $q_{\text{max}}=40, p=0.5$, which yield a Bloom filter length of approximately $23905$\,bits.

According to Figure~\ref{fig:bloomDistance} and thus the number of changed bits for valid genomic sequences and artificial queries, an arbitrary threshold can be defined.
For example, if the necessary Hamming distance between submitted Bloom filters is set to $50$\,bits, consecutive changes of up to $31$ characters or random changes of up to $2$ characters are detected.
This would roughly equal the performance of what we showed in section~\ref{sec:goodRand} for a Hamming distance $\delta'$ of $32$.

This gap between valid and artificial queries can be increased even further by increasing the value $q_{\text{max}}$ within the VGRAM algorithm.

\section{Conclusion}
\label{sec:conclusion}

In this paper we investigated how to mitigate inference attacks on privacy-preserving genome matching.
Since randomization approaches, such as differential privacy, cannot work, we employ a detection technique using an error-correcting code and a one-way function.
A novel zero-knowledge proof ensures honesty on part of the querier.
This zero-knowledge proof is more efficient than generic techniques by using somewhat homomorphic encryption and also more efficient than directly proofing the error-correction by introducing the codeword.
We show an analysis that our scheme (in combination with limiting the number of queries) still effectively detects and mitigates Goodrich's inference attack.


\begin{thebibliography}{10}

\bibitem{AtaKer03}
M.~Atallah, F.~Kerschbaum, and W.~Du.
Secure and Private Sequence Comparisons.
{\em Proceedings of the ACM Workshop on the Privacy in Electronic Society (WPES)},
2003.

\bibitem{AtaLi04}
M.~Atallah, and J.~Li.
Secure Outsourcing of Sequence Comparisons.
{\em Proceedings of the 4th International Workshop on Privacy Enhancing Technologies (PET)},
2004.

\bibitem{BalBar11}
P.~Baldi, R.~Baronio, E.~De Cristofaro, P.~Gasti, and G.~Tsudik.
Countering GATTACA: Efficient and Secure Testing of Fully-Sequenced Human Genomes.
{\em Proceedings of the 18th ACM Conference on Computer and Communications Security (CCS)},
2011.

\bibitem{BecKer12}
M.~Beck, and F.~Kerschbaum.
Approximate Two-Party Privacy-Preserving String Matching with Linear Complexity.
{\em Proceedings of the 2nd IEEE International Congress on Big Data (BIGDATA)},
2013.

\bibitem{BelChe04}
S. Bellovin, and W. Cheswick.
Privacy-Enhanced Searches Using Encrypted Bloom Filters.
{\em Columbia University Technical Report CUCS-034-07},
2004.

\bibitem{BlaAli10}
M.~Blanton, and M.~Aliasgari.
Secure Outsourcing of DNA Searching via Finite Automata.
{\em Proceedings of the 24th IFIP Conference on Data and Applications Security and Privacy (DBSEC)},
2010.

\bibitem{BlaAta12}
M.~Blanton, M.~Atallah, K.~Frikken, and Q.~Malluhi.
Secure and Efficient Outsourcing of Sequence Comparisons.
{\em Proceedings of the 17th European Symposium on Research in Computer Security (ESORICS)},
2012.

\bibitem{BonGoh05}
D.~Boneh, E.~Goh, and K.~Nissim.
Evaluating 2-DNF Formulas on Ciphertexts.
{\em Proceedings of the 2nd Theory of Cryptography Conference (TCC)},
2005.


\bibitem{CamMic99}
J.~Camenisch, and M.~Michels.
Proving in Zero-Knowledge that a Number is the Product of Two Safe Primes.
{\em Advances in Cryptology (EUROCRYPT)},
1999.

\bibitem{ChePen12}
Y.~Chen, B.~Peng, X.~Wang, and H.~Tang.
Large-Scale Privacy-Preserving Mappings of Human Genomic Sequences on Hybrid Clouds.
{\em Proceedings of the 19th Network and Distributed System Security Symposium (NDSS)},
2012.

\bibitem{DamJur01}
I.~Damg{\aa}rd, and M.~Jurik.
A Generalisation, a Simplification and Some Applications of Paillier's Probabilistic Public-Key System.
{\em Proceedings of 4th International Workshop on Practice and Theory in Public Key Cryptography (PKC)},
2001.

\bibitem{DavFra98}
G.~Davida, Y.~Frankel, and B.~Matt.
On Enabling Secure Applications Through Off-Line Biometric Identification.
{\em Proceedings of the 19th IEEE Symposium on Security and Privacy},
1998.

\bibitem{Dom08}
J.~Domingo-Ferrer.
A Survey of Inference Control Methods for Privacy-Preserving Data Mining.
In Privacy-Preserving Data Mining, Eds.~C.~Aggarwal, and P.~Yu.
{\em Advances in Database Systems 34}, 2008.

\bibitem{DurXue12}
E.~Durham, Y.~Xue, M.~Kantarcioglu, and B.~Malin.
Quantifying the Correctness, Computational Complexity, and Security of Privacy-Preserving String Comparators for Record Linkage.
{\em Information Fusion 13(4)}, 2012.

\bibitem{Dwo08}
C.~Dwork.
Differential Privacy: A Survey of Results.
{\em Proceedings of the 5th International Conference on Theory and Applications of Models of Computation (TAMC)}, 
2008.

\bibitem{FraKat11}
M.~Franz, S.~Katzenbeisser, B.~Deiseroth, K.~Hamacher, S.~Jha, and H.~Schr\"oder.
Towards Secure Bioinformatics Services.
{\em Proceedings of the 15th International Conference on Financial Cryptography and Data Security (FC)},
2011.

\bibitem{Gol04}
O.~Goldreich.
Foundations of Cryptography: Volume 2 -- Basic Applications.
{\em Cambridge University Press},
2004.

\bibitem{Goo09}
M.~Goodrich.
The Mastermind Attack on Genomic Data.
{\em Proceedings of the 30th IEEE Symposium on Security and Privacy},
2009.

\bibitem{Gra01}
L.~Gravano, P.~Ipeirotis, H.~Jagadish, N.~Koudas, S.~Muthukrishnan, and D.~Srivastava.
Approximate String Joins in a Database (Almost) for Free.
{\em Proceedings of the 27th International Conference on Very Large Data Bases (VLDB)}, 2001.

\bibitem{Gro10}
J.~Groth.
A Verifiable Secret Shuffe of Homomorphic Encryptions.
{\em Journal of Cryptology 23(4)},
2010.

\bibitem{HuaEva11}
Y.~Huang, D.~Evans, J.~Katz, and L.~Malka.
Faster Secure Two-Party Computation Using Garbled Circuits.
{\em Proceedings of the 20th USENIX Security Symposium},
2011.

\bibitem{mtDnaDb06}
M.~Ingman, and U.~Gyllensten.
mtDB: Human Mitochondrial Genome Database, a Resource for Population Genetics and Medical Sciences.
{\em Nucleic Acids Research 34}, 2006.

\bibitem{JhaKru08}
S.~Jha, L.~Kruger, and V.~Shmatikov.
Towards Practical Privacy for Genomic Computation.
{\em Proceedings of the 29th IEEE Symposium on Security and Privacy},
2008.

\bibitem{JueWat99}
A.~Juels, and M.~Wattenberg.
A Fuzzy Commitment Scheme.
{\em Proceedings of the 6th ACM Conference on Computer and Communications Security (CCS)},
1999.

\bibitem{KanJia08}
M.~Kantarcioglu, W.~Jiang, Y.~Liu, and B.~Malin.
A Cryptographic Approach to Securely Share and Query Genomic Sequences.
{\em IEEE Transactions on Information Technology in Biomedicine 12(5)},
2008.

\bibitem{KuzKan11}
M.~Kuzu, M.~Kantarcioglu, E.~Durham, and B.~Malin.
A Constraint Satisfaction Cryptanalysis of Bloom Filters in Private Record Linkage.
{\em Proceedings of the 11th International Symposium on Privacy Enhancing Technologies (PETS)},
2011.

\bibitem{Li07}
C.~Li, B.~Wang, and X.~Yang.
VGRAM: Improving Performance of Approximate Queries on String Collections using Variable-Length Grams.
{\em Proceedings of the 33rd International Conference on Very Large Databases (VLDB)}, 2007.

\bibitem{McS09}
F.~McSherry.
Privacy Integrated Queries.
{\em Proceedings of the ACM International Conference on Management of Data (SIGMOD)},
2009.

\bibitem{Mul54}
D.~Muller.
Application of Boolean Algebra to Switching Circuit Design and to Error Detection. 
{\em IRE Transactions on Electronic Computers 3},
1954.

\bibitem{PapSib10}
O.~Papapetrou, W.~Siberski, and W.~Nejdl.
Cardinality Estimation and Dynamic Length Adaptation for Bloom Filters.
{\em Distributed and Parallel Databases 28(2-3)},
2010.

\bibitem{Ree54}
I.~Reed.
A Class of Multiple-Error-Correcting Codes and the Decoding Scheme.
{\em Transactions of the IRE Professional Group on Information Theory 4},
1954.

\bibitem{TroKat07}
J.~Troncoso-Pastoriza, S.~Katzenbeisser, and M.~Celik.
Privacy-Preserving Error-Resilient DNA Searching Through Oblivious Automata.
{\em Proceedings of the 14th ACM Conference on Computer and Communications Security (CCS)},
2007.

\bibitem{WanLi09}
R.~Wang, Y.~Li, X.~Wang, H.~Tang, and X.~Zhou.
Learning Your Identity and Disease from Research Papers: Information Leaks in Genome Wide Association Study.
{\em Proceedings of the 16th ACM Conference on Computer and Communications Security (CCS)},
2009.


\bibitem{Yao86}
A.~Yao.
How to Generate and Exchange Secrets. 
{\em Proceedings of the 27th IEEE Foundations of Computer Science (FOCS)},
1986.

\end{thebibliography}
\end{document}